\documentclass[11pt,a4paper]{article}

\usepackage{graphicx}
\usepackage{amsmath,amssymb,amsfonts}
\usepackage{newtxtext,newtxmath}
\usepackage[T1]{fontenc}
\usepackage{microtype}
\usepackage{silence}
\usepackage{url}
\usepackage{cite}
\usepackage{geometry}
\usepackage{setspace}
\usepackage{titlesec}
\usepackage{caption}
\usepackage{fancyhdr}
\usepackage{etoolbox}

\titleformat{\section}[hang]{\normalfont\fontsize{12}{14.5}\selectfont\bfseries}{\thesection}{0.65em}{}
\titlespacing*{\section}{0pt}{2.1ex plus .6ex minus .2ex}{.9ex plus .2ex}
\titleformat{\subsection}[hang]{\normalfont\fontsize{11}{13.5}\selectfont\itshape\bfseries}{\thesubsection}{0.65em}{}
\titlespacing*{\subsection}{0pt}{1.5ex plus .4ex minus .2ex}{.55ex plus .1ex}
\titleformat{name=\section,numberless}[hang]{\normalfont\fontsize{11}{13.5}\selectfont\bfseries}{}{0pt}{}
\titlespacing*{name=\section,numberless}{0pt}{2ex}{.8ex}

\renewenvironment{abstract}
{\par\vspace{.7em}\noindent{\normalfont\bfseries\fontsize{10.5}{12.5}\selectfont Abstract}\par\vspace{.45em}\begingroup\small\setstretch{1.17}\noindent\ignorespaces}
{\par\endgroup\vspace{1.25em}}

\begin{document}

\title{\vspace{-1.5em}\normalfont\fontsize{16}{19}\selectfont\bfseries Symmetry-Engineered Moiré Metasurfaces for Multibeam Control\vspace{-.35em}}

\author{%
Yu Chen\textsuperscript{1,*},
Lan Ke\textsuperscript{1,*},
Shaojie Ma\textsuperscript{2,\dag},
Shuo Liu\textsuperscript{1,\S}\\[0.5em]
\small $^{1}$State Key Laboratory of Millimeter Waves, Southeast University, Nanjing 210096, China\\
\small $^{2}$College of Future Information Technology, Fudan University, Shanghai 200433, China\\[0.4em]
\small\textsuperscript{*}These authors contributed equally to this work.\\
\small Corresponding authors:
\textsuperscript{\ensuremath{\dagger}}shaojiema@fudan.edu.cn;
\small\textsuperscript{\S}liushuo.china@seu.edu.cn
}

\date{\vspace{-2.0em}}
\maketitle\vspace{-1.2em}

\begin{abstract}
Moiré metasurfaces offer a versatile platform for wavefront engineering through the relative rotation of periodic structures. Despite extensive demonstrations of continuous beam steering, the physical mechanisms governing beam multiplicity remain poorly understood, hindering the systematic design of multibeam radiation states. Here, we establish a predictive framework for engineering beam multiplicity in reflective moiré metasurfaces based on the dominant Fourier coefficients of their constituent patterns. We reveal that the rotational symmetry of these Fourier components in reciprocal space fully determines the symmetry of the resulting moiré unit cell and, consequently, its radiation characteristics. By further incorporating their phase information, both beam multiplicity and steering trajectories can be accurately predicted. We demonstrate this approach using reflective moiré metasurfaces constructed from patterns with p6m, p3m1, and p4m plane-group symmetries, achieving one-, two-, three-, four-, and six-beam radiation states with continuously tunable steering trajectories. Beyond symmetry-based design, we identify two additional degrees of freedom for beam multiplicity control. Selecting different rotation centers reshapes the phase distributions of the dominant Fourier coefficients, enabling further manipulation of beam multiplicity and spatial arrangement, while varying the relative twist angle alters the pairing of Fourier components, inducing transitions between distinct beam multiplicities. These findings establish a systematic framework for versatile and reconfigurable multibeam wavefront engineering.
\end{abstract}

\section{Introduction}\label{sec1}

Moiré patterns arise from the superposition of two periodic structures with a relative rotational misalignment \cite{RN1}. First explored in twisted electronic materials, moiré superlattices have become a powerful platform for engineering flat electronic bands and realizing exotic quantum states such as superconductivity \cite{RN2,RN3,RN4,RN5}. The concept of moiré engineering has rapidly extended beyond electronic systems to a broad range of wave systems, encompassing optics \cite{RN6,RN7,RN8,RN9,RN10,RN11}, acoustics \cite{RN12,RN13,RN14,RN15}, microwave \cite{RN16,RN17,RN18} and thermal systems \cite{RN19,RN20}. The development of moiré systems has provided a versatile platform for exploring and tailoring a broad range of wave phenomena, such as bound states in the continuum \cite{RN21,RN22,RN23,RN24}, optical vortices \cite{RN25,RN26,RN27}, high-quality-factor lasers \cite{RN28,RN29}, and moiré flat bands \cite{RN30,RN31,RN32,RN33}. 

Beyond their ability to engineer wave confinement and localization, moiré metasurfaces have emerged as a versatile platform for beam manipulation and steering. By cascading two metasurface layers with a controlled relative rotation, the generated beam can be dynamically reconfigured in both propagation direction and focal position. This moiré-based strategy has enabled diverse beam-control functionalities, including axial varifocal focusing \cite{RN34,RN35,RN36}, lateral focal-spot steering \cite{RN37,RN38,RN39}, dynamic terahertz beam control \cite{RN40,RN41}, holographic beam shaping \cite{RN42,RN43}, polarization-multiplexed vectorial wavefront manipulation \cite{RN38,RN44,RN45} and polarization-selective filtering \cite{RN46}. Moiré beamforming metasurfaces enable reconfigurable wave control solely by adjusting the relative rotation between two patterned layers, substantially reducing implementation complexity and cost while enhancing system robustness \cite{RN47,RN48}. These advantages make them an attractive candidate for next-generation wireless communication systems and a potential alternative to reconfigurable intelligent surfaces (RISs) \cite{RN49,RN50}.

In 2022, we initially proposed a reflective moiré metasurface for steering the reflected wave to symmetric directions \cite{RN51}. Distinct from previous moiré metasurfaces based on either thick dielectric structures or cascaded configurations, our reflective moiré metasurface employs a zero-spacing moiré pattern formed by two metallic layers, thereby effectively reducing the overall device thickness. However, similar to other reported studies on moiré metasurfaces \cite{RN15,RN52,RN53}, our previously proposed reflective moiré metasurfaces were restricted to four- and six-beam radiation patterns, limiting their adaptability in communication and sensing applications. In addition, previous studies have primarily focused on predicting beam trajectories through reciprocal-space vector analysis \cite{RN51,RN52,RN53,RN54,RN55}, whereas the physical principles governing beam multiplicity have received little attention. More specifically, the relationship between different combinations of periodic patterns and the resulting beam multiplicity remains unclear.

Here, we establish a predictive framework for engineering beam multiplicity in reflective moiré metasurfaces based on Fourier-coefficient phase synthesis. We show that the phase distributions of moiré components, synthesized from the dominant Fourier components of constituent patterns, determine the number of radiated beams. Using patterns with p6m, p3m1, and p4m plane-group symmetries, we realize one-, two-, three-, four-, and six-beam radiation states, covering all rotational orders allowed by crystallographic symmetry in two-dimensional periodic systems. Beyond symmetry-based design, we identify two additional degrees of freedom for beam multiplicity control: rotation-center selection and relative twist-angle adjustment. Selecting different rotation centers alters the phases of the dominant Fourier coefficients, thereby reshaping the synthesized phase distributions and resulting radiation states. In contrast, varying the relative twist angle changes the pairing of Fourier components, enabling transitions between distinct beam multiplicities. The resulting far-field radiation patterns agree well with theoretical predictions. This framework establishes a direct link between Fourier-coefficient phases, moiré structural symmetry, and radiation characteristics, providing a systematic approach to predictive and reconfigurable multibeam engineering while reducing reliance on computationally intensive full-wave simulations.

\section{Results}\label{sec2}

\subsection{Reflective moiré metasurface and symmetry-dependent beam multiplicity}\label{subsec2.1}

Figure 1a illustrates the operating principle of the proposed reflective moiré metasurface. Superimposing two periodic metasurface patterns with a relative rotation generates a moiré pattern that enables dynamic manipulation of the reflected wavefront. By adjusting the relative twist angle, the reflected beams can be continuously steered in elevation, providing a mechanically reconfigurable beamforming mechanism. As shown in Figure 1b, the proposed metasurface consists of two independently rotatable metasurfaces, a low-index spacer, and a metallic back reflector. Detailed structural parameters and fabrication procedures are provided in Section 4.1.

\begin{figure}[htbp]
\centering
\includegraphics{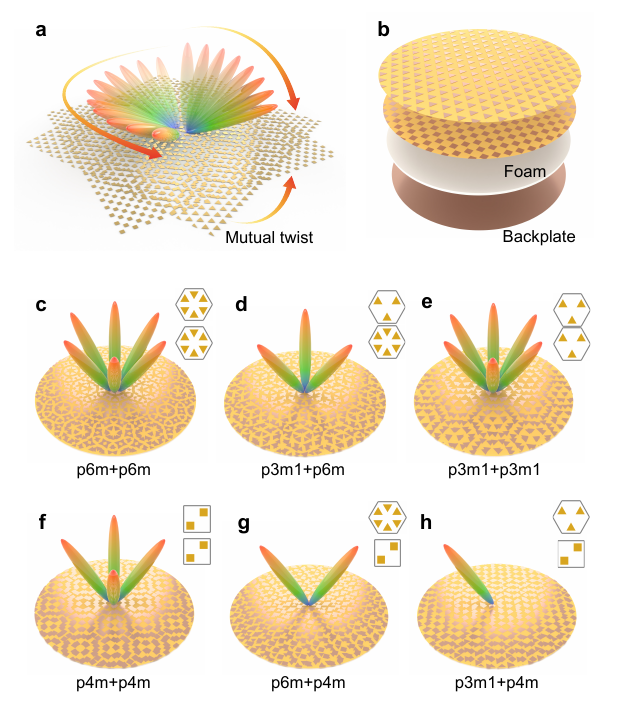}
\caption{\textbf{Schematic illustration of moiré metasurfaces.}
\textbf{a,} Dynamic beam manipulation enabled by the relative rotation of two periodic patterns with arbitrary rotational symmetries.
\textbf{b,} Structural configuration of a reflective moiré metasurface.
\textbf{c–h,} Moiré metasurfaces formed by different combinations of pattern symmetries and their corresponding radiation patterns:
\textbf{c,} p6m + p6m, six beams;
\textbf{d,} p3m1 + p6m, three beams;
\textbf{e,} p3m1 + p3m1, six beams;
\textbf{f,} p4m + p4m, four beams;
\textbf{g,} p6m + p4m, two beams;
\textbf{h,} p3m1 + p4m, a single beam.}
\label{fig1}
\end{figure}

Beyond continuous beam steering, the moiré platform offers additional control over beam multiplicity through symmetry engineering of the constituent patterns. Different combinations of rotational symmetries give rise to distinct moiré structures and corresponding radiation states. Figures 1c–h present six representative configurations constructed from patterns with p6m, p3m1, and p4m plane-group symmetries, together with their far-field radiation patterns. The p6m + p6m, p3m1 + p6m, p3m1 + p3m1, p4m + p4m, p6m + p4m, and p3m1 + p4m combinations generate six-, three-, six-, four-, two-, and single-beam radiation, respectively. These examples demonstrate the versatility of symmetry engineering in tailoring beam multiplicity and spatial radiation characteristics. As this work focuses on rotational symmetry, mirror and glide symmetries are not considered independently. Accordingly, our analysis is restricted to plane groups containing mirror symmetry.

\subsection{Predicting beam multiplicity from the phase of moiré components}\label{subsec2.2}

To establish a general rule governing beam multiplicity in moiré metasurfaces, we examine the reciprocal-space characteristics of the constituent patterns and their moiré combinations (see Methods Section 4.2 for details). Figures 2a–c show the unit cells of patterns with p6m, p3m1, and p4m plane-group symmetries, together with the phases of their dominant Fourier coefficients. These coefficients correspond to reciprocal-lattice indices (1,0), (1,1), (0,1), (-1,0), (-1,-1), and (0,-1) for the p6m and p3m1 patterns, and (1,1), (-1,1), (1,-1), and (-1,-1) for the p4m pattern. The resulting phase distributions reflect the rotational symmetries of the corresponding unit cells (Supplementary Information 1). We then derive the moiré components in reciprocal space by convolving the dominant Fourier components of the constituent patterns. Owing to their discrete Fourier spectra, this convolution reduces to pairwise addition of the dominant reciprocal-lattice vectors, with the associated Fourier coefficients combined to determine the synthesized phases (Supplementary Information 2). As illustrated in Figures 2d–i, different combinations of pattern symmetries generate distinct sets of moiré components in reciprocal space. The rotational symmetry of these components identifies groups of equivalent radiation channels, which can be selectively enhanced or suppressed by combining patterns with different symmetries, thereby controlling the number of radiated beams.

\begin{figure}[htbp]
\centering
\includegraphics{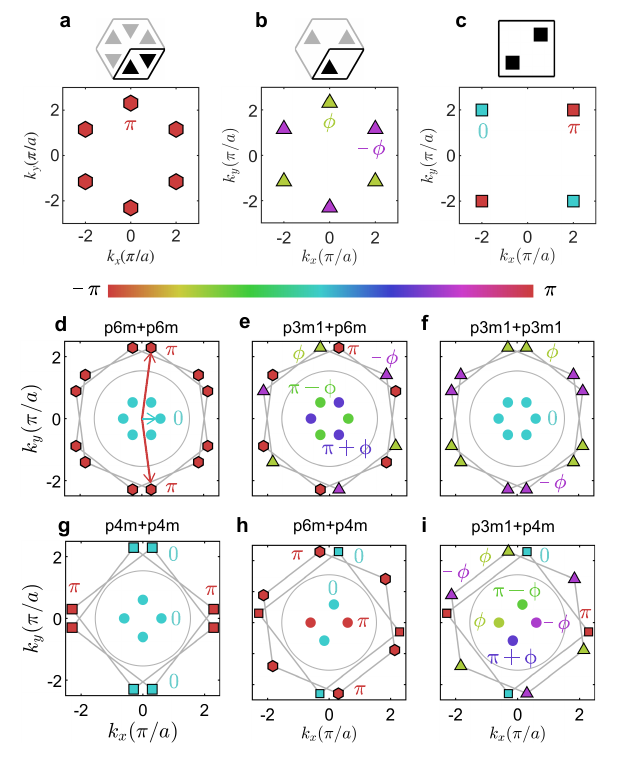}
\caption{\textbf{Principle of Fourier-coefficient phase synthesis.}
\textbf{a–c,} Unit cells of periodic patterns with different plane-group symmetries—\textbf{a,} p6m; \textbf{b,} p3m1; and \textbf{c,} p4m—together with the phases of their dominant Fourier coefficients in reciprocal space.
\textbf{d–i,} Fourier-coefficient phase-synthesis analysis for six pairwise combinations of these patterns: \textbf{d,} p6m + p6m; \textbf{e,} p3m1 + p6m; \textbf{f,} p3m1 + p3m1; \textbf{g,} p4m + p4m; \textbf{h,} p6m + p4m; and \textbf{i,} p3m1 + p4m.}
\label{fig2}
\end{figure}

For the p6m pattern, the sixfold rotational symmetry of the unit cell ensures that the six dominant first-order diffraction components are related by 60° rotations and possess identical amplitudes and phases. When two p6m patterns are combined (p6m + p6m; Figure 2d), all contributing pairs of diffraction components yield the same synthesized phase. Consequently, six dominant moiré components with equal amplitudes and phases emerge within the light cone, giving rise to six-beam radiation. Replacing one layer with a p3m1 pattern preserves the six dominant diffraction components but introduces a threefold-symmetric phase distribution, with phases alternating between $\varphi$ and $-\varphi$. In the p3m1 + p6m configuration (Figure 2e), these components combine with those of the p6m layer to produce two distinct synthesized phases, $\pi-\varphi$ and $\pi+\varphi$. The resulting moiré components therefore exhibit threefold rotational symmetry, forming two interleaved triplets with alternating phases. By optimizing the pattern parameters, one triplet can be efficiently excited, yielding three-beam radiation, while the other triplet is simultaneously suppressed. When both layers adopt the p3m1 pattern (p3m1 + p3m1; Figure 2f), each contains diffraction components with two distinct phases. However, in the configuration considered here, each contributing pair consists of components with opposite phases, $\varphi$ and $-\varphi$, resulting in a common synthesized phase for all six moiré components and restoring six-beam radiation.

The p4m pattern supports four dominant diffraction components with phases alternating between $0$ and $\pi$. In the p4m + p4m configuration (Figure 2g), the contributing component pairs fall into two groups, yielding synthesized phases of $0+0=0$ and $\pi+\pi=2\pi\equiv0\pmod{2\pi}$, respectively. Thus, all four dominant moiré components share a common phase, giving rise to four-beam radiation. For the p4m + p6m configuration (Figure 2h), the different lattice geometries produce a more complex distribution of moiré components. The two groups with the smallest in-plane wavevector magnitudes exhibit uniform phases of $0$ and $\pi$, respectively. By appropriately optimizing the structural parameters, one group can be preferentially excited, producing a symmetric two-beam radiation pattern. Replacing the p6m layer with a p3m1 pattern (p3m1 + p4m; Figure 2i) breaks the phase equivalence within these groups, resulting in four moiré components with distinct phases. With suitable structural parameters, a single radiation channel can be efficiently excited, yielding single-beam radiation. Detailed reciprocal-space vector constructions and corresponding schematic illustrations for these six configurations are provided in Supplementary Figure S7.

These results demonstrate that the reciprocal-space arrangement of the synthesized Fourier components determines the available radiation channels, while their phase distributions govern which channels are efficiently excited and, consequently, the resulting beam multiplicity. By analyzing only the dominant Fourier components of the constituent patterns, both beam multiplicity and radiation directions can be predicted without relying on computationally intensive full-wave simulations. Moreover, the beam directions can be continuously tuned by varying the relative twist angle between the two patterns. The beam multiplicity remains unchanged as long as the pairing relationships between the dominant Fourier components, and hence the resulting moiré phase, are preserved.

\subsection{Numerical simulations and experimental verification}\label{subsec2.3}

To validate the Fourier-coefficient phase-synthesis method, we performed full-wave numerical simulations of the proposed moiré metasurfaces using the commercial electromagnetic simulation software CST. The structure was illuminated by a left-handed circularly polarized (LHCP) plane wave, and the reflected LHCP component was extracted for analysis. To evaluate the radiation efficiency, all far-field amplitudes were normalized to the maximum reflected radiation intensity of a perfect electric conductor (PEC) plate with the same aperture size. The geometric parameters of the patterns were chosen such that the dominant moiré components lie within the light cone (Supplementary Information 3).

Representative simulation results for four different combinations are shown in Figures 3a-d, while the remaining cases are provided in Supplementary Information 3. Consistent with the Fourier-coefficient phase-synthesis predictions, different symmetry combinations give rise to various beam multiplicities, such as one-, two-, three-, four- and six-beam radiation states. Furthermore, varying the relative twist angle continuously modifies the elevation angles of the radiated beams. Theoretical beam trajectories predicted by synthesizing Fourier components are compared with the simulated radiation directions in Figures 3e-3h. Excellent agreement is observed over the entire tuning range of twist angles. Although several weak reflection beams deviate from the theoretical beam trajectories, the principal radiation directions closely follow the theoretical predictions. These results validate the Fourier-coefficient phase-synthesis method for accurately predicting both beam multiplicity and beam angle.

\begin{figure}[htbp]
\centering
\includegraphics[
  width=\textwidth,
  height=0.9\textheight,
  keepaspectratio
]{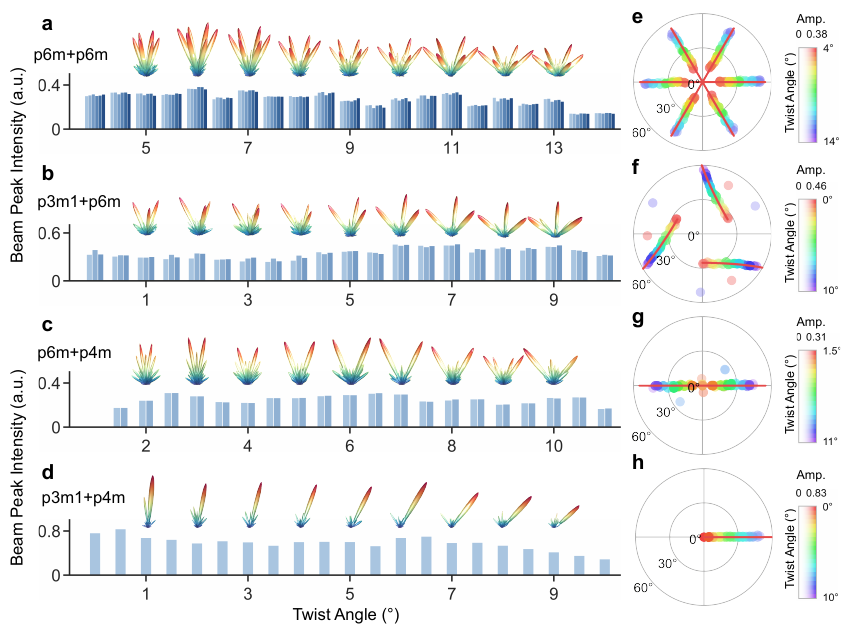}
\caption{\textbf{Numerically simulated far-field radiation characteristics of the p6m + p6m, p3m1 + p6m, p6m + p4m, and p3m1 + p4m pattern combinations at different twist angles.}
\textbf{a-d,} Three-dimensional far-field radiation patterns and corresponding bar plots of the normalized intensity of individual beams at different twist angles.
\textbf{e-h,} Comparison of beam angles predicted by the Fourier-coefficient phase-synthesis method (red lines) with those obtained from full-wave simulations (colorful dots).}
\label{fig3}
\end{figure}

To experimentally validate the proposed Fourier-coefficient phase-synthesis method, moiré metasurfaces with different pattern symmetry combinations were fabricated and characterized through far-field measurements in a microwave anechoic chamber. The samples consist of two closely stacked metasurface layers, a low-permittivity foam layer, and an all-metallic reflecting ground plane. The two metasurfaces were mounted in a rotatable fixture to control the relative twist angle. We measured the cross-polarized far-field response using a pair of circularly polarized horn antennas and a vector network analyzer (Ceyear 3674G). Details of the fabrication and measurement procedures are provided in the Section 4.1. Figure 4 shows the measured radiation patterns for four symmetry combinations: p6m + p6m, p3m1 + p6m, p6m + p4m, and p3m1 + p4m. The remaining cases are presented in Supplementary Information 3. The experimental results are in excellent agreement with both the theoretical predictions and full-wave simulations. Distinct one-, two-, three-, four-, and six-beam radiation states are experimentally demonstrated, together with continuous tuning of the beam elevation angle over the twist-angle range considered in the theoretical analysis. These results validate the Fourier-coefficient phase-synthesis method as an accurate and reliable approach for predicting and controlling both beam multiplicity and beam direction in moiré metasurfaces.

\begin{figure}[htbp]
\centering
\includegraphics[
  width=\textwidth,
  height=0.9\textheight,
  keepaspectratio
]{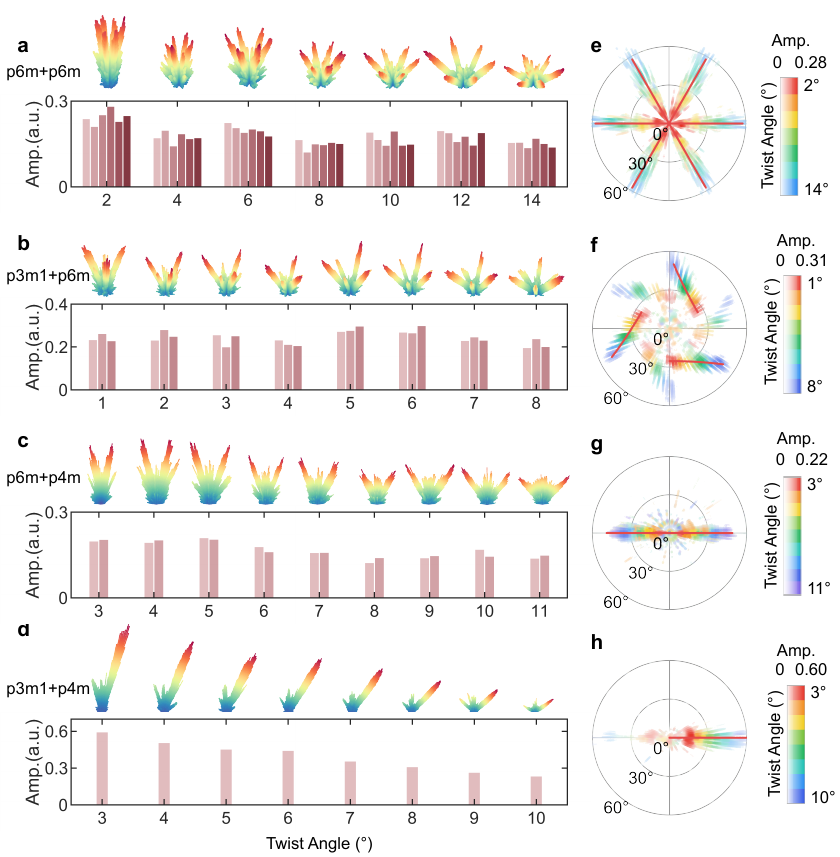}
\caption{\textbf{Experimentally measured far-field radiation characteristics of the p6m + p6m, p3m1 + p6m, p6m + p4m, and p3m1 + p4m pattern combinations at different twist angles.}
\textbf{a–d,} Measured three-dimensional far-field radiation patterns and corresponding bar plots of the normalized intensity of individual beams at different twist angles. Colors distinguish different relative twist angles, while transparency indicates the normalized beam intensity.
\textbf{e–h,} Comparison of beam trajectories predicted by the Fourier-coefficient phase-synthesis method (red lines) with experimentally measured trajectories (red dots) for the four pattern combinations, respectively.}
\label{fig4}
\end{figure}

Interestingly, the measurements reveal an additional effect that cannot be captured by the idealized  numerical simulation, in which the two metallic patterns are in direct contact (zero separation). In the fabricated samples, the two metallic patterns are separated by a 5-$\mu m$-thick polyimide layer to facilitate relative rotation. This finite separation modifies the electromagnetic coupling between the two metallic patterns, leading to a slight deviation from the ideal radiation pattern. As shown in Figures 4b and 4f, weak neighboring beams appear on the opposite side of the principal beam in the asymmetric three-beam radiation pattern. Numerical simulations show that the radiation pattern becomes noticeably affected when the interlayer separation exceeds 0.03 mm and 0.06 mm for the one-beam and three-beam cases, respectively (Supplementary Information 4). These results suggest that particular care should be taken during sample assembly to minimize the interlayer separation and avoid such effects.

\subsection{Beam multiplicity control by varying the rotation center or relative twist angle}\label{subsec2.4}

The preceding results establish a predictive relationship between the rotational symmetries of constituent patterns and beam multiplicity through Fourier-coefficient phase synthesis. This section further identifies two additional degrees of freedom for controlling beam multiplicity: rotation-center selection and relative twist-angle adjustment. Specifically, shifting the rotation center alters the phases of the dominant Fourier components, thereby reshaping the synthesized phase distribution and modifying the number of radiated beams (Figure 5). By contrast, varying the relative twist angle changes how these Fourier components are paired during moiré synthesis, enabling beam multiplicity control without altering the underlying phase distributions (Figure 6). Below, we examine these two mechanisms in detail.

For a periodic pattern, selecting a different rotation center is equivalent to an in-plane translation, which alters the phases of its Fourier coefficients. The resulting phase shifts can be analytically determined (Supplementary Information 5). Figure 5a illustrates rotation centers of different symmetry orders within the unit cells of the p6m and p4m patterns. Choosing one of these points as the rotation center of the moiré metasurface is equivalent to translating the corresponding patterns in the opposite direction relative to the rotation axis. Consequently, the rotational symmetry associated with each selected center is reflected in the resulting Fourier-coefficient phase distribution. Figure 5b shows representative  phase distributions of the dominant Fourier components for the three rotation centers indicated in Figure 5a. For the p6m pattern, selecting different centers yields additional phase distributions with threefold and twofold rotational symmetries, whereas the p4m pattern yields an additional distribution with fourfold rotational symmetry. Combining patterns with different rotation centers produces distinct synthesized phase distributions, giving rise to far-field radiation patterns with different beam multiplicities and spatial arrangements. Representative configurations are presented in Figures 5c–h, with the complete set of combinations and corresponding parameters provided in Supplementary Information 5.

For the p4m + p4m combination, selecting twofold rotation centers in both layers results in four identical synthesized components (Figure 2g), thereby producing four radiation beams. When a twofold rotation center is selected in one layer and a fourfold rotation center in the other, the dominant synthesized components exhibit twofold rotational symmetry (Figure 5c). These components can be divided into two groups with phases of 0 and $\pi$, respectively. By tuning the structural parameters, either group can be selectively enhanced to achieve efficient radiation, resulting in two beams distributed along the $y$-direction or the $x$-direction, respectively. In contrast, when fourfold rotation centers are selected in both layers (Figure 5d), all four synthesized components exhibit zero phase, enabling four-beam radiation.

\begin{figure}[htbp]
\centering
\includegraphics{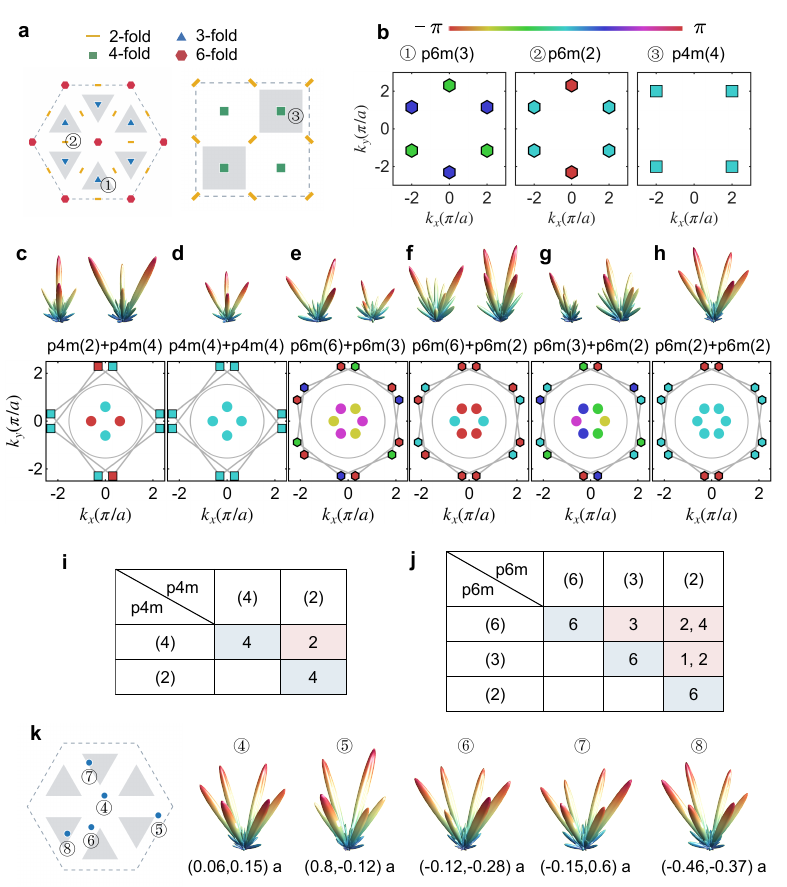}
\caption{\textbf{Control of beam multiplicity by varying the rotation center.}
\textbf{a,} Schematic illustration of the rotation centers within the p6m and p4m unit cells.
\textbf{b,} Phase distributions of the dominant Fourier coefficients obtained from the Fourier transforms of patterns defined by the rotation centers indicated in (a). Numbers in parentheses denote the rotational symmetry order at each center.
\textbf{c–h,} Schematic illustrations of moiré-component synthesis using patterns with different rotation centers, together with the corresponding simulated three-dimensional far-field radiation patterns.
\textbf{i–j,} Summary of beam multiplicities for different rotation-center combinations in \textbf{i,} p4m + p4m and \textbf{j,} p6m + p6m configurations. Blue shading indicates configurations supporting continuous beam steering with a fixed beam multiplicity, whereas red shading denotes those for which no such operating regime has been identified.
\textbf{k,} Representative rotation centers selected away from high-symmetry positions within the p6m unit cell and the corresponding simulated three-dimensional far-field radiation patterns.}
\label{fig5}
\end{figure}

The p6m + p6m configuration exhibits similar behavior but supports a richer variety of synthesized phase combinations, giving rise to diverse radiation patterns, including three-beam radiation (Figure 5e), two beams on opposite sides and four-beam radiation (Figure 5f), single-beam radiation and two beams on the same side (Figure 5g), and six-beam radiation (Figure 5h). Notably, by tuning the structural parameters, different radiation channels can be selectively enhanced, enabling switching between three-beam patterns with different orientations (Figure 5e), between oppositely directed two-beam and four-beam radiation (Figure 5f), and between single-beam and same-side two-beam radiation (Figure 5g).

Figures 5i and 5j summarize the beam multiplicities obtained for different rotation-center combinations in the two systems. Blue indicates configurations that support continuous beam steering with a fixed beam multiplicity, whereas red denotes those for which no such operating regime has been identified. For configurations with matched rotation-center symmetries, the symmetry-related phase relationships among the dominant moiré components are preserved during relative twisting, stabilizing the radiation channels throughout beam scanning. In contrast, mismatched rotation-center symmetries can introduce chiral moiré configurations with reduced symmetry, making the excitation of different radiation channels more sensitive to the twist angle and thus complicating the maintenance of a fixed beam multiplicity under unchanged structural parameters. Beyond the differences in steering behavior, both tables reveal a consistent trend: selecting a common rotation center for the two layers yields the maximum beam multiplicity. This behavior can be understood from Fourier-coefficient phase synthesis, as identical translations of the two patterns preserve equal phases among the dominant synthesized moiré components (Supplementary Information 5). For the p6m + p6m configuration, all six dominant components consequently retain zero synthesized phase, supporting six-beam radiation. To further verify this prediction, we selected several arbitrary rotation centers away from high-symmetry positions within the p6m unit cell. As shown in Figure 5k, six distinct radiation beams are consistently obtained in all cases.

\begin{figure}[htbp]
\centering
\includegraphics{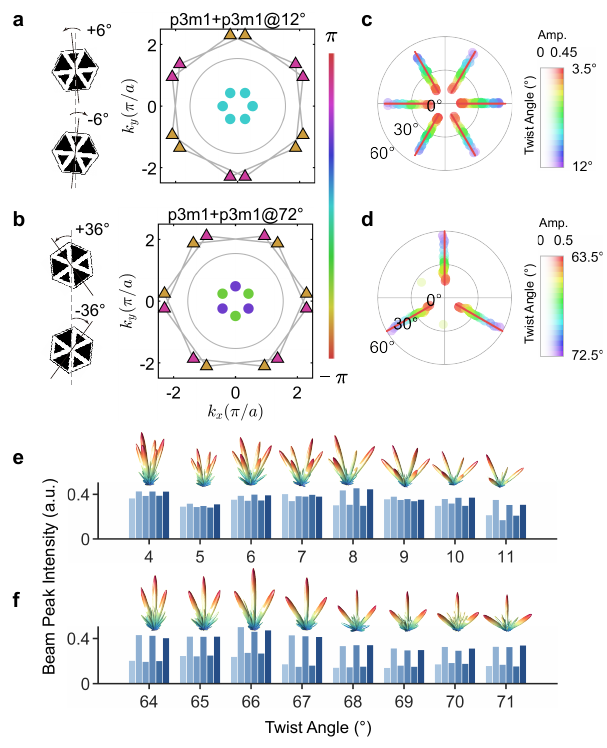}
\caption{\textbf{Control of beam multiplicity through relative rotation.}
\textbf{a–b,} Fourier-coefficient phase-synthesis analysis of the p3m1 + p3m1 combination at relative twist angles of \textbf{a,} 12$^\circ$ and \textbf{b,} 72$^\circ$.
\textbf{c–d,} Comparison of beam directions predicted by the Fourier-coefficient phase-synthesis method (red lines) with those obtained from full-wave simulations (colorful dots) for the configurations in (a) and (b), respectively.
\textbf{e–f,} Full-wave-simulated three-dimensional far-field radiation patterns and corresponding beam-intensity distributions for the configurations in (a) and (b), respectively.}
\label{fig6}
\end{figure}

The preceding examples demonstrate beam control through varying the rotation center, which combines distinct phase distributions of Fourier coefficients. Relative rotation offers a complementary mechanism by altering the pairing of Fourier components during moiré synthesis, thereby enabling access to different radiation states. To illustrate this mechanism, we consider the p3m1 + p3m1 configuration, employing an alternative p3m1 pattern with an expanded phase-tuning range to improve radiation performance (Supplementary Information 5). For twist angles below 30$^\circ$, all six moiré components share the same phase, giving rise to six-beam radiation (Figs. 6a and 6c). In contrast, when the twist angle exceeds 60$^\circ$, these components separate into two groups with distinct phases, reproducing the phase configuration of the p3m1 + p6m combination. As a result, only three radiation channels remain efficiently excited, driving a transition from six-beam to three-beam radiation (Figures 6b–6d). Figures 6e and 6f present the radiation patterns and corresponding beam amplitudes at different twist angles, confirming that relative rotation enables dynamic control over both beam direction and multiplicity. This additional degree of freedom substantially enhances the flexibility of the proposed approach for beamforming applications.

\section{Discussion}\label{sec3}
In this work, we establish a Fourier-coefficient phase-synthesis method for analyzing and engineering beam radiation in moiré metasurfaces. We show that the phase distribution of the synthesized moiré components in reciprocal space governs the radiation characteristics. By considering only the dominant Fourier components, the beam multiplicity, elevation angle, and azimuth angle can be predicted without full-wave numerical simulations. Based on this method, we demonstrate multibeam radiation control by combining patterns with different rotational symmetries, realizing one-, two-, three-, four-, and six-beam radiation states with continuously tunable beam directions. Both numerical simulations and experimental measurements confirm the predicted beam multiplicities and far-field radiation performance. The observed sensitivity to interlayer spacing suggests that it may provide an additional degree of freedom for beam multiplicity engineering. Furthermore, we demonstrate that selecting different rotation centers provides an effective approach to controlling beam multiplicity by modifying the phase distributions of the dominant Fourier coefficients and, consequently, the synthesized moiré components. In addition, varying the relative twist angle changes the pairing of these Fourier components, allowing dynamic switching between different beam multiplicities while simultaneously controlling the beam directions.

Overall, the proposed Fourier-coefficient phase-synthesis method offers a new strategy for engineering beam multiplicity through combinations of periodic patterns with different rotational symmetries. The additional degrees of freedom introduced by rotation center selection and relative twist angle adjustment further expand the range of accessible radiation states, providing a flexible route toward multibeam control and wavefront shaping across microwave, terahertz, infrared, and even visible frequencies. Beyond the reflective moiré metasurfaces demonstrated here, the method can potentially be extended to transmissive and coaxial-fed radiating configurations, offering a versatile platform for reconfigurable beamforming devices.

\section{Methods}\label{sec4}
\subsection{Fabrication and measurement}\label{subsec4.1}
The reflective moiré metasurface comprises two independently fabricated metallic metasurface layers, a dielectric spacer, and a metallic ground plane. The metallic patterns were fabricated on ultrathin dielectric substrates with a thickness of 0.05 mm and a copper thickness of 0.035 mm. A 3-mm-thick low-permittivity foam spacer was inserted between the metasurface layers and the ground plane to establish the reflective configuration. The dielectric substrates have a relative permittivity of 3.5 and a loss tangent of 0.03, whereas the foam spacer has a relative permittivity of 1.1 and a loss tangent of 0.001. The metasurface features a circular aperture with a diameter of 250 mm, ensuring a constant effective aperture during relative rotation.

Experimental characterization was performed in a microwave anechoic chamber. The metasurface layers were fabricated on flexible dielectric substrates using standard printed circuit board (PCB) technology (Supplementary Information 4). After adjusting the relative twist angle, the two layers were assembled in a custom-designed fixture and mounted on a horizontal wooden platform. A circularly polarized horn antenna was used to illuminate the metasurface with a left-handed circularly polarized (LHCP) wave, approximating plane-wave incidence. A second circularly polarized horn antenna was positioned in the far-field region to receive the right-handed circularly polarized (RHCP) reflected wave. The scattering parameters (S-parameters) were measured using a vector network analyzer connected to both antennas, from which the three-dimensional far-field radiation patterns were reconstructed.

\subsection{Fourier expansion of a two-dimensional periodic pattern}\label{subsec4.2}
The Fourier coefficients of a periodic pattern are defined as
\[
F\left( \mathbf{G}_{mn} \right) =\frac{1}{\left| \Omega \right|}\int_{\Omega}f\left(\mathbf{r}\right)e^{-i\mathbf{G}_{mn}\cdot\mathbf{r}}d^2\mathbf{r},
\tag{1}
\]
where $\Omega$ denotes the unit cell domain and $\mathbf{G}_{mn}=m\mathbf{b}_1+n\mathbf{b}_2$ is a reciprocal lattice vector, with $\mathbf{b}_1$ and $\mathbf{b}_2$ being the reciprocal lattice basis vectors. For the hexagonal patterns considered here, $f(\mathbf{r})$ is piecewise constant, taking values of $\Phi_1$ and $\Phi_2$ in the metallic and air regions, denoted by $\Omega_1$ and $\Omega_2$, respectively. The Fourier coefficients can therefore be expressed as
\[
F\left( \mathbf{G}_{mn} \right) =
\frac{1}{\left| \Omega \right|}\int_{\Omega_1}\Phi_1e^{-i\mathbf{G}_{mn}\cdot\mathbf{r}}d^2\mathbf{r}
+\frac{1}{\left| \Omega \right|}\int_{\Omega_2}\Phi_2e^{-i\mathbf{G}_{mn}\cdot\mathbf{r}}d^2\mathbf{r}.
\tag{2}
\]
Introducing the geometric integral
\[
S_j\left(\mathbf{G}_{mn}\right)=\int_{\Omega_j}e^{-i\mathbf{G}_{mn}\cdot\mathbf{r}}d^2\mathbf{r},\quad j=1,2,
\tag{3}
\]
yields
\[
F\left(\mathbf{G}_{mn}\right)=
\frac{\Phi_1S_1\left(\mathbf{G}_{mn}\right)+\Phi_2S_2\left(\mathbf{G}_{mn}\right)}{\left|\Omega\right|}.
\tag{4}
\]
Since $\Omega_1$ and $\Omega_2$ together constitute the entire unit cell, their geometric integrals satisfy $S_1(\mathbf{G}_{mn})+S_2(\mathbf{G}_{mn})=0$ for all nonzero diffraction orders, i.e., $(m,n)\neq(0,0)$. Equation (4) thus simplifies to
\[
F\left(\mathbf{G}_{mn}\right)=
\frac{\Phi_1-\Phi_2}{\left|\Omega\right|}
S_1\left(\mathbf{G}_{mn}\right),\quad (m,n)\neq(0,0).
\tag{5}
\]
This expression shows that, apart from a constant phase offset associated with $\Phi_1-\Phi_2$, the phases of the nonzero-order Fourier coefficients are governed by the geometry and spatial arrangement of the metallic regions within the unit cell. For the hexagonal patterns, the Fourier coefficients can be evaluated analytically by dividing the unit cell into six equal-area triangles and integrating over their respective domains. The amplitudes and phases of different diffraction orders are then obtained by substituting the corresponding integer indices $m$ and $n$. The same procedure applies to the p4m pattern shown in Figure 2c, with appropriate modifications to the reciprocal lattice vectors and integration domain.

\end{document}